\documentclass[fleqn,showkeys,twocolumn,amssymb,eps,12pt]{revtex4-1}
\usepackage{graphicx,bm,amssymb,amsfonts,hyperref,upgreek,wrapfig}
\usepackage[rightcaption]{sidecap}
\usepackage{graphics,epstopdf,ifpdf}
\usepackage{subscript}
\usepackage{subcaption}
\usepackage{float}
\usepackage{blindtext}
\usepackage{nicefrac}
\begin{document}
\title [Tunnel and barrier time-delay and the weak measurement limit]
{Universal Behavior of Tunneling Time and  Barrier 
Time-Delay Decoupling in Attoclock Measurements}
 \author{Ossama Kullie} 
 \affiliation{\scriptsize \tiny Theoretical Physics, Institute for Physics, 
 Department of Mathematics and Natural Science, University of Kassel, 
 Germany}
 \thanks{\tiny Electronic mail: kullie@uni-kassel.de}
\sffamily 
\begin{abstract}
\scriptsize
 The measurement of the tunneling time-delay is hotly debated and remains
 controversial.  
 In previous works, we showed that a model that accurately describes the 
 time-delay measured by the attoclock experiment in adiabatic and 
 nonadiabatic field calibrations.
 In the present work, we show that the tunneling time reveals a universal 
 behavior with disentangled contributions. 
 Even more remarkable is that the barrier tunneling time-delay can be 
 convincingly defined and determined from the difference between the 
 time-delay of adiabatic and nonadiabatic tunnel-ionization, which also 
 show good agreement with experimental results.
 Furthermore, we illustrate that in the weak measurement limit, the barrier
 time-delay corresponds to the Larmor-clock time and the interaction 
 time within the barrier.
 \end{abstract}
 \keywords{\scriptsize Ultrafast science, attosecond physics, tunneling 
 and tunnel ionization time-delay, nonadiabatic effects, weak measurement, 
 Larmor clock and interaction time.} 
 \maketitle     
 \scriptsize 
\section{Theoretical model}\label{ssec:int}  
\begin{figure}[h]
\vspace{-0.25cm}
\centering
 \includegraphics[width=15.0cm,height=10.cm]{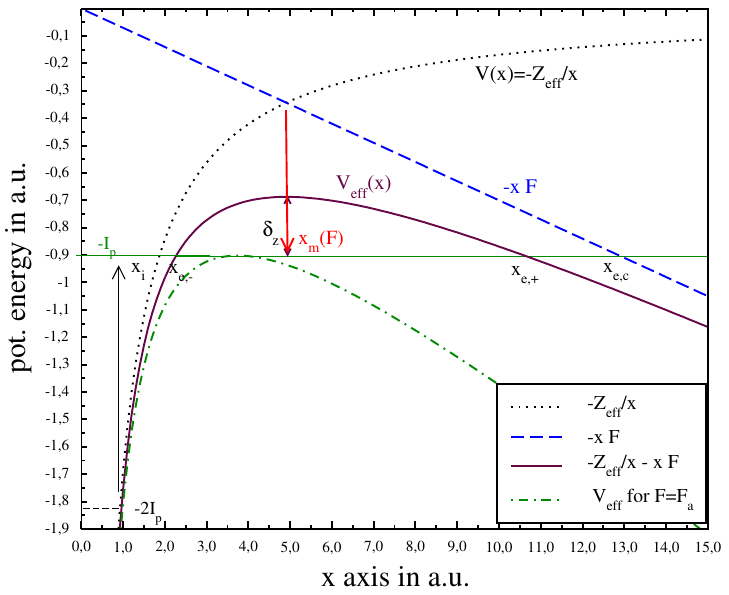}
\hspace{-5.50cm}
\vspace{-6.0cm}
 \caption{\label{figptc}\scriptsize (Color online) A sketch of the 
 atomic and the effective potential curves, showing the inner 
 $x_{\rm e,-}$ and outer $x_{\rm e,+}$ points, the width and  height of the 
 barrier formed by the laser field.  
 $x_{\rm e,\pm} = \nicefrac{(I_{\rm p} \pm \delta_z)}{2F}$,
 {\scriptsize$d_{\rm B}= x_{\rm e,+}-x_{\rm e,-}=\nicefrac{\delta_z}{F}$}, 
 {\tiny $\delta_z=\sqrt{I_{\rm p}^{2}-4 Z_{\rm eff} F}$}.
 The plot is for He-atom in the SAEA model with 
 $Z_{\rm eff}=1.6875$ and $I_{\rm p}\approx0.9\, au$ 
 \cite{Kullie:2015,Kullie:2024}.
 For different systems the overall picture remains the same.
 }
\end{figure}

 The interaction of an atom with a laser pulse in strong-field and 
 attosecond science can be modeled in a simplified manner 
 \cite{Kullie:2015,Kullie:2020,Kullie:2024} as shown in figure \ref{figptc}. 
 In \cite{Kullie:2015} a simple tunneling model is developed with 
 expressions involving basic laser and atomic parameters that describe  
 the measurement result of the attoclock.
 This model provides insight into the temporal properties of tunneling 
 ionization and sheds light on the role of time in quantum mechanics. 
 We briefly present our model below, in which an electron can be 
 tunnel-ionized by a laser pulse with an electric field strength 
 (hereafter field strength) $F$. 
 Throughout this work $F$ stands for {\it the peak electric field strength 
 at maximum} (quasistatic approximation \cite{Popruzhenko:2022}), 
 and atomic units ($au$) are used ($\hbar = m = e = 1$). 
 A direct ionization happens when the field strength reaches a threshold  
 called atomic field strength $F_{a}=\nicefrac{I_{\rm p}^{2}}{(4 Z_{\rm eff})}$  
 \cite{Augst:1989,Augst:1991}, where $I_{\rm p}$ is the ionization potential 
 of the system (atom or molecule) and $Z_{\rm eff}$ is the effective nuclear 
 charge in the single-active electron approximation (SAEA).
 However, for $F<F_{a}$ the ionization can happen by tunneling through 
 an effective potential barrier including the Coulomb 
 potential of the core and the electric field of the laser pulse. 
 It can be expressed in a one-dimensional form in the length gauge due 
 to  the G\"oppert-Mayer gauge-transformation  \cite{Goeppert:1931} by
 \begin{equation}\label{efpot}
  V_{\rm eff}(x)=V(x)-x F =-\frac{Z_{\rm eff}}{x}-x F,
 \end{equation}
 compare figure \ref{figptc}. 
 In the model the tunneling process can be described solely by the 
 ionization potential $I_{\rm p}$ of the valence (interacting) electron 
 and the peak field strength $F$, where the barrier height is given by 
 {$\delta_z=\mbox{\scriptsize $\sqrt{I_{\rm p}^{2}-4 Z_{\rm eff} F}$}$}.  
 In figure  \ref{figptc} (for details see \cite{Kullie:2015}), the inner 
 $x_{\rm e,-}$ and outer $x_{\rm e,+}$ points are given by  
 $x_{\rm e,\pm} = \nicefrac{(I_{\rm p} \pm \delta_z)}{2F}$.
 The inner and outer points are called entrance and exit points, 
 respectively.
 The barrier width is {\scriptsize $d_{\rm B}= x_{\rm e,+}-x_{\rm e,-}
 =\nicefrac{\delta_z}{F}$}, 
 and its (maximum) height (at {\scriptsize $x_{m}(F)=
 \sqrt{\nicefrac{Z_{\rm eff}}{F}}$}) 
 is $\delta_z$. 
 At $F=F_{a}$ we have $\delta_z=0$ ($d_{\rm B}=0$), i.e., the barrier 
 disappears and the direct (barrier-suppression) ionization begins, 
 (green (dashed-doted) curve in figure \ref{figptc}).
\subsection{\scriptsize \bf Adiabatic tunneling}\label{pa:adt}
 In the adiabatic field calibration by Landsman et. al. 
 \cite{Landsman:2014II}, we showed in \cite{Kullie:2015} that the 
 tunneling time-delay can be expressed by the following forms,
 \begin{eqnarray}\label{Tdi}
 &&\tau_{_{\rm T,d}}=\frac{1}{2(I_{\rm p}-\delta_z)}, \quad 
 \tau_{_{\rm T,i}}=\frac{1}{2(I_{\rm p}+\delta_z)} 
 \end{eqnarray}
 In \cite{Kullie:2015} was also shown that $\tau_{_{\rm T,d}}$ in eq \ref{Tdi} 
 agrees well with the experimental result of Landsman et. al. 
 \cite{Landsman:2014II}. 
 Their physical reasoning is the following:
 $\tau_{_{\rm T,d}}$ is the time-delay of the adiabatic tunnel-ionization,   
 with respect to the ionization at atomic field strength $F_{a}$, 
 required to overcome the barrier and escape the atom into the continuum 
 \cite{Kullie:2015}.
 Furthermore, $\tau_{_{\rm T,i}}$ is the time needed to reach the entrance  
 point $x_{_{e,-}}$ from the initial point $x_i$, compare figure 
 \ref{figptc}. 
 At the limit of atomic field strength,  we have 
 $\lim\limits_{F\to F_{a}} \delta_z= 0$, 
 $\lim\limits_{F\to F_{a}} \tau_{_{\rm T,d}}= 
 \frac{1}{2I_{\rm p}}=\tau_{a}$ (see below).  
 For $F> F_{a}$, the barrier-suppression ionization sets on
 \cite{Delone:1998,Kiyan:1991}.
 On the opposite side, we have $\lim\limits_{F\to 0} \delta_z= I_{\rm p}$, 
 $\lim\limits_{F\to 0} \tau_{_{\rm T,d}}= \infty$, so nothing happens and 
 the electron remains in its ground state undisturbed, which shows that 
 our model is consistent. 
 For details, see \cite{Kullie:2020,Kullie:2015,Kullie:2016,Kullie:2018}.
\subsection{\scriptsize \bf Nonadiabatic tunneling}
 In the nonadiabatic field calibration of Hofmann et. al. 
 \cite{Hofmann:2019}, we found in \cite{Kullie:2024} that the 
 time-delay of the nonadiabatic tunnel-ionization is descried by the 
 relation
 \begin{eqnarray}\label{tdion}
 \tau_{\rm dion}(F) = 
\frac{1}{2} \frac{I_{\rm p}}{4Z_{\rm eff} F}&=&
\frac{1}{2I_{\rm p}} \frac{I_{\rm p}^{2}}{4Z_{\rm eff} F}\\
&=&
 \frac{1}{2I_{\rm p}}\frac{F_{a}}{F}=\tau_a\, \xi({F})
 \end{eqnarray}
 In \cite{Kullie:2024} it was shown that Eq. \ref{tdion} agrees well with 
 the experimental result of Hofmann et. al. \cite{Hofmann:2019}. 
 In addition, the result was confirmed by the numerical integration of  
 the time-dependent Schr\"odinger equation (NITDSE) \cite{Kullie:2024}.   
 It is seen that in the limit $F\to F_{a}, \tau_{\rm dion}=\tau_a$, 
 whereas in the opposite case $F\to 0, \tau_{\rm dion}=\infty$, similar 
 to $\tau_{_{\rm T,d}}$.
 Hence, $\tau_a$ is always (adiabatic and nonadiabatic) a lower quantum  
 limit of the tunnel-ionization time-delay and  does not quantum 
 mechanically vanish. 
 $\xi(F)$ is an enhancement factor for field strength $F<F_{a}$.
\section{The barrier time-delay}\label{ssec:tt}
 We have seen in the previous section that the experimental results 
 of the tunneling time-delay, in the two field calibrations (adiabatic 
 \cite{Landsman:2014II}, nonadiabatic \cite{Hofmann:2019}), are in 
 good agreement with our tunneling model, as given by the adiabatic 
 and nonadiabatic tunneling time-delay $\tau_{_{\rm T,d}}$ and 
 $\tau_{\rm dion}$ of eqs \ref{Tdi}, \ref{tdion} respectively. 
   
 In this section, we show that the time-delay due to the barrier itself, 
 which can be considered as the actual tunneling time-delay, can be 
 decoupled and determined by considering the adiabatic and nonadiabatic 
 field calibrations together.
 
 The time-delay due to the barrier itself is, in fact, the time-delay 
 due to crossing the barrier region.
 However, the experiment does not verify whether the time spent in the 
 barrier is affected by some reflections of the electrons inside the 
 barrier before they escape the atom (see further below Eq. \ref{tdelt}). 
 Apart from that, multiple reflections are related the regime of high 
 harmonics generation, where electrons recombine by emitting
 high harmonics.
 In the attoclock experiment, the momentum distribution of photoelectrons 
 (ions) during the tunnel-ionization process is considered to extract and 
 determine the delay-time \cite{Eckle:2008,Eckle:2008s}.  
 
 Eq. \ref{Tdi} can be decomposed into a twofold time-delay with respect 
 to ionization at $F_{a}$, representing $\tau_{_{\rm T,d}}$ in an unfolded 
 form, 
\begin{eqnarray}\label{TdF}
\tau_{_{\rm T,d}}&=&\frac{1}{2(I_{\rm p}-\delta_z)}=  
\frac{1}{2}\frac{I_{\rm p}}{4Z_{\rm eff}F} 
 \left(1+\frac{\delta_z}{I_{\rm p}}\right)\\\nonumber 
 &=&\frac{1}{2I_{\rm p}}\frac{F_{a}}{F}\left(1+ 
 \frac{\delta_z}{I_{\rm p}}\right) 
 =\tau_a (\xi(F) +  \Lambda(F)) \\\nonumber
 &=&\tau_{\rm dion}+\tau_{\rm dB}\equiv \tau_{\rm Ad}
\end{eqnarray}
 Below we refer to $\tau_{\rm Ad}\equiv \tau_{_{\rm T,d}}$ as adiabatic 
 tunnel-ionization. 
 From the second line of Eq. \ref{TdF} follows that the adiabatic time-delay 
 can be interpreted as a time-delay with respect to ionization at atomic 
 field strength with $\tau_a=\tau_{{\rm Ad}}(F_{a})=\nicefrac{1}{(2I_{\rm p})}$,    
  where both terms ($\tau_{\rm dion}, \tau_{{\rm dB}}$) present real time-delay. 
 Again $(\xi({F}) + \Lambda({F}))$ is an enhancement factor for 
 field strength $F<F_{a}$.
 
 The first term in the last line of Eq. \ref{TdF}, $\tau_{\rm dion}$,  
 which is given in Eq. \ref{tdion}, is a time-delay solely because 
 $F$ is smaller than $F_{a}$ and saturates at $F=F_a$, whereas the 
 second term $\tau_{\rm dB}$ is a time-delay due to the barrier itself 
 and thus, can be considered as the actual tunneling time-delay
 \cite{Kullie:2020}.
 We will show that it can be decoupled from the first term and is thus 
 the tunneling time solely to the presence of the barrier itself.
 This can be seen from the factor ${\delta_z}/{I_{\rm p}}$, which relates 
 the barrier height for $F<F_a$ to the ionization potential, which in  
 turn represents the maximum barrier height  
 ($\lim\limits_{F\to 0}\delta_z=I_{\rm p}$). 
 It vanishes when the barrier $\lim\limits_{F\to 0}\delta_{z}=0$ 
 vanishes, i.e. when saturation is reached at $F=F_a$.
 Indeed, Winful \cite{Winful:2003} and Lunardi \cite{Lunardi:2019} 
 proposed expressions that resemble the adiabatic tunnel-ionization 
 time-delay given in Eq. \ref{TdF}.
 \begin{eqnarray}\label{Win}
  \tau^{\rm Winf}_g&=&\tau_{\rm si}+\tau_{\rm dwell}\\\label{Lun}
  \tau^{\rm Lun}_{T}&=&\tau_{\rm well}+\tau_{\rm barrier} 
\end{eqnarray}
 In the quantum tunneling of a wave packet or a flux of particles 
 scattering on a potential barrier, Winful showed that the group 
 time-delay or the Wigner time-delay can be written in the form of 
 Eq. \ref{Win}, where $\tau_{\rm dwell}$ is the dwell time which 
 corresponds to our $\tau_{\rm dB}$, and $\tau_{\rm si}$ is according 
 to Winful a self-interference term which corresponds to our 
 $\tau_{\rm dion}$.  
 Importantly, Winful showed that the contribution of the first term 
 is disentangled from the barrier time-delay \cite{Winful:2003}.
 In their work Lunardi et. al. used the so-called  
 Salecker-Wigner-Peres quantum-clock (SWP-QC) with MC-simulation 
 \cite{Lunardi:2018,Lunardi:2019}.
 They found that the tunneling time in the attoclock is given by the 
 form of Eq. \ref{Lun}. 
  \begin{figure}[t]
  \vspace{-1cm}
  \centering 
  \resizebox{7.0cm}{9.cm}{\includegraphics{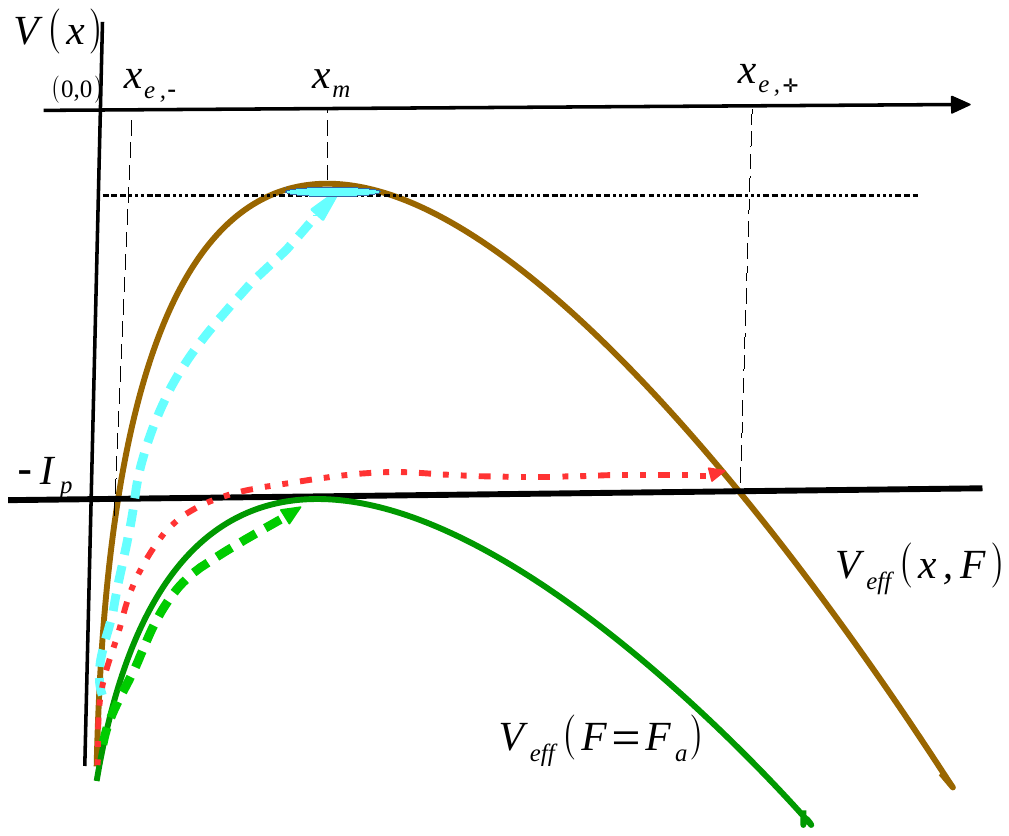}}
\vspace{-4.cm}
 \caption{\label{figTI} \scriptsize
  Illustration of 
 tunnel-ionization of the  horizontal (dashed-dotted, red) and vertical 
 (dashed, blue) channels. The dotted line is a virtual state, as well 
 as the  BSI (green).}
\end{figure}
 The first term, which they called well-time is, according to the 
 authors, the time (-delay) that elapses before reaching the barrier.  
 This corresponds to our $\tau_{\rm dion}$.
 The second term, $\tau_{\rm barrier}$, corresponds to the barrier time 
 (-delay) which corresponds to our $\tau_{\rm dB}$. 
 The three results (ours confirmed with the NITDSE, Winful's and 
 Lunardi's) clearly show that the separation of Eq. \ref{TdF} presents  
 a unified T-time picture (UTTP) \cite{Winful:2003}, which allows us 
 to conclude that the tunneling time-delay due the barrier itself is 
 given by $\tau_{\rm dB}= \tau_{\rm dwell}=\tau_{\rm barrier}$ and can 
 be determined from eqs. \ref{tdion}, \ref{TdF} as follows:    
 \begin{equation}\label{tdelt}
 \tau_{_{\rm Ad}}-\tau_{\rm dion}=\tau_{\rm dB}=\frac{1}{2}
 \frac{1}{4Z_{\rm eff}} \frac{\delta_z}{F}=\frac{1}{2}
 \frac{d_{\rm B}}{4Z_{\rm eff}}
 \end{equation} 
 In Eq. \ref{tdelt}, $\tau_{\rm dB}$ is the barrier (tunneling) time-delay, 
 usually referred to as the time spent within the barrier,  
 $\tau_{\rm barrier}$ by Lunardi or the dwell time $\tau_{\rm dwell}$ 
 by Winful.
 One also finds a linear dependence of $\tau_{\rm dB}$ on the barrier 
 width $d_{\rm B}$. 
 The barrier time-delay tends to infinity 
 as $F\to 0$, 
 leaving the stationary ground state of the atom undisturbed.
 The barrier time-delay tends to zero in the limit of $F\to F_{a} 
 \, (\delta_z=d_{\rm B}=0)$ because the barrier disappears, as is known from  
 the seminal work of Hartman \cite{Hartman:1962}, where the linear 
 dependence on the barrier width in Eq. \ref{tdelt} indicates that 
 multiple reflections (if they exist) have a negligible 
 influence on the tunnel-ionization. 
 Obviously, $\tau_{\rm dB}$ cannot be measured by the experiment directly  
 since the first term is always present, $\tau_{\rm dion}, \tau_{\rm si}, 
 \tau_{\rm well}$.
 However, as seen in Eq. \ref{tdelt}, it can be determined taking both 
 field calibrations into account, thus decoupling the barrier 
 (tunneling) time-delay $\tau_{\rm dB}$ in the tunnel-ionization of   
 the attoclock experiment.    

 It is worth noting that the decomposition given in Eq. \ref{TdF} was derived
 in \cite{Kullie:2020} concerning controversial issue of a quantum operator. 
 However, its significance emerged when we found in \cite{Kullie:2024} that 
 the first term $\tau_{\rm dion}$ agrees well with the experimental result 
 of Hofmann \cite{Hofmann:2019} in nonadiabatic field calibration and the 
 numerical integration of time-dependent Schr\"odinger equation.  
 Furthermore, the relevance of numerical results of Winful and Lunardi 
 in Eqs. \ref{Win}-\ref{Lun} was not clear for experimental finding of 
 the attoclock. 
 Our model and the present work are of crucial importance as we argue 
 for a conceivable method to define and determine the barrier 
 time delay, which is a hotly debated topic.
 
 An illustration of the tunnel-ionization of adiabatic (horizontal) and 
 nonadiabatic (vertical) channels is shown in figure \ref{figTI}.
 Fortunately, for the He atom we can consider the results of Landsman 
 \cite{Landsman:2014II} (adiabatic calibration) and Hofmann  
 \cite{Hofmann:2019} (nonadiabatic calibration) since they belong to 
 the same experiment \cite{Eckle:2008,Eckle:2008s}. 
 In figure \ref{figtdelt} (a), we show both experimental results with 
 curves obtained by data fitting.
 The curves show a $\nicefrac{1}{F}$-dependence as expected, 
 subtracting them from each other gives the experimental barrier time, 
 which is denoted by $\tau_{\rm barrier,exp}$. 
 
 In figure \ref{figtdelt} (b), we compare the barrier (tunneling) 
 time-delay $\tau_{\rm dB}$ of Eq. \ref{tdelt} with the experimental 
 counterpart $\tau_{\rm barrier,exp}$ (the orange curve).
 We plot $\tau_{\rm dB}$ for two $Z_{\rm eff}=1.0, 1.344=
 \mbox{\scriptsize $\sqrt{2I_{\rm p}}$}$, expanding the range to a larger 
 barrier width (smaller $F$) than specified in the experimental data.
 As can be seen in figure \ref{figtdelt} (b), the agreement is very good.
 It undoubtedly shows that the time spent in the barrier (the 
 time-delay caused by the barrier upon tunnel-ionization in attosecond 
 experiment) is $\tau_{\rm dB}$, which can be determined 
 (from the experimental data) by $\tau_{\rm barrier,exp}$.
 The agreement (with $\tau_{\rm barrier,exp}$ the orange curve) becomes 
 better for $Z_{\rm eff}=1$ ($Z_{1}$ green curve) in the region of small field 
 strength, since the barrier width $d_{\rm B}=\nicefrac{\delta_{z}}{F}$ 
 is large and the tunnel-ionized electron escapes far from the atomic core. 
 For larger field strength (near the atomic field strength) the  
 barrier width is small and the agreement is better for 
 {\scriptsize $Z_{\rm eff}= \sqrt{2I_{\rm p}}=1.344$} ($Z_{2}$ blue curve). 
 Finally, all curves tend to zero for $F\rightarrow F_{a}$, since the 
 barrier disappears as expected.
 
 In figure \ref{figtdelt} (b) we additionally present a curve 
 ($\tau_{\rm LC}$) of Larmor clock (LC) time, which was also obtained 
 by fitting of the LC data given by Hofmann \cite{Hofmann:2019}.  
 The agreement with LC (magenta curve) is better for small $F$ values, 
 which is the limit of thick barrier (more on this in the next section). 
 This is because Hofmann's LC time-delay is given for the nonadiabatic 
 field calibration and our $\tau_{\rm dB}$ becomes closer to nonadiabatic 
 time-delay $\tau_{\rm dion}$ for small $F$ (thick barrier) (see Eq.  
 \ref{tTdelt} below).
 This again explains the disagreement for larger 
 field strength $F=0.06-0.1$, compare figure \ref{figtdelt} (b). 
 The notation $\tau_{\rm dB}$ below is used to refer to the barrier 
 (tunneling) time-delay $\tau_{\rm dB}\, (\equiv \tau_{\rm barrier}
 \equiv\tau_{\rm dwell})$ with the corresponding adiabatic and 
 nonadiabatic tunnel-ionization  time-delay $\tau_{\rm Ad}$, 
 $\tau _{\rm dion}$, respectively.  
\section{Weak measurement limit} \label{ssec:wm}
 \begin{figure}[t]
 \vspace*{0.50cm}
 \centering
 \resizebox{7.25cm}{5.cm}{{\large\textbf{(a)}}\includegraphics{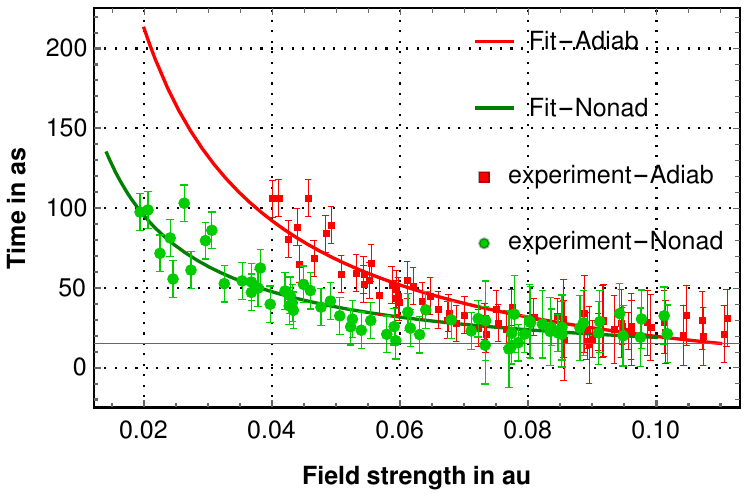}}
 \resizebox{7.5cm}{5.0cm}{{\large\textbf{(b)}}\includegraphics{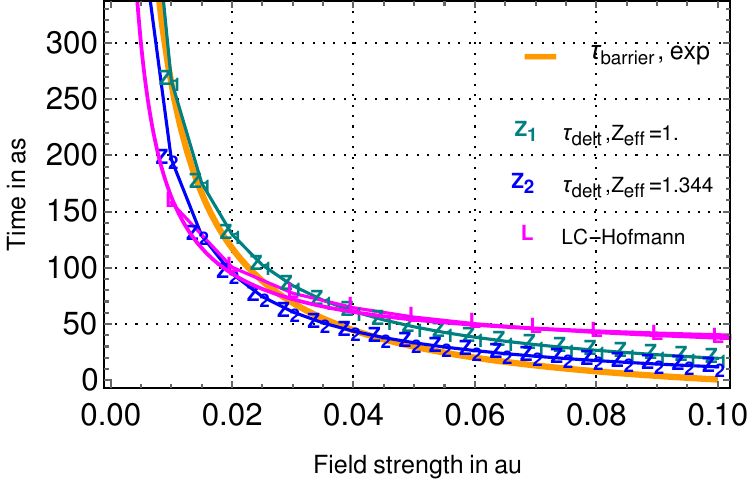}}
  \caption{\label{figtdelt} \scriptsize
 (a): tunnel-ionization time-delay versus field strength.  
 Experimental data of adiabatic \cite{Landsman:2014II} (red) and 
 nonadiabatic \cite{Hofmann:2019} (green) field calibrations with their 
 data-fit curves. 
 (b): barrier time-delay $\tau_{\rm barrier,exp}$ (orange) obtained 
 by subtractions of the data-fit curves (shown left), and the barrier 
 time-delay $\tau_{\rm dB}$ of Eq. \ref{tdelt} for $Z_{\rm eff}=1,1.344$ 
 (green, blue).
 As well as the LC-curve (magenta) obtained by data-fit of LC-data 
 given in  \cite{Hofmann:2019}.}
 \vspace{-0.25cm}
 \end{figure}
 In the previous section we found that the LC time $\tau_{\rm LC}$ 
 tends closer to the barrier time-delay $\tau_{\rm dB},\, 
 \tau_{\rm barrier,exp}$ for small field strength, compare figure 
 \ref{figtdelt} (b). 
 The LC-time is usually considered in the context of the so-called 
 weak measurement (WM) approach, which characterizes a system before 
 and after it interacts with a measurement apparatus 
 \cite{Aharonov:1988,Rozema:2012}.
 Landsman \cite{Landsman:2015} showed that the WM value of 
 the time-delay corresponds to LC-time $\tau_{\rm weak}=\tau_{\rm LC}$ 
 \cite{Landsman:2015,Steinberg:1995}. 
 Thus, for small field strengths ($F\ll F_{a}$) the barrier time-delay 
 $\tau_{\rm dB}$ in our model corresponds to the WM approach, which is 
 clearly seen in figure \ref{figtdelt} (b).
 Indeed, for $F\ll F_{a}$ the barrier width $d_{\rm B}=\nicefrac{\delta_z}{F}$ 
 becomes large enough which makes the barrier thick . 
 At the limit of thick or opaque barrier, $d_{\rm B}$ approaches the so-called 
 classical barrier width $d_{\rm C}$, $\lim\limits_{F\ll F_{a}} 
 d_{\rm B}\approx \nicefrac{I_{\rm p}}{F}=d_{\rm C}$ 
 ($\equiv x_{\rm e,c}$), compare figure \ref{figptc}. 
 In this limit the barrier height 
 $\lim\limits_{F\ll F_{a}} {\delta_z= I_{\rm p}}$, and the barrier time 
 \begin{equation}\label{tTdelt}
 \lim\limits_{F\ll F_{a}}\tau_{\rm dB}=\tau_{a}
 \frac{F_{a}}{F} \frac{(\delta_z\approx I_{\rm p})}{I_{\rm p}}
 \approx  \tau_{a} \frac{F_{a}}{F}=\tau_{\rm dion}
 \end{equation}
 This result is interesting, as it shows that the barrier time-delay 
 $\tau_{\rm dB}$ for a thick (or opaque) barrier is approximately equal 
 to tunnel-ionization time-delay $\tau_{\rm dion}$ (Eq. \ref{tdion}) of 
 the nonadiabatic field calibration \cite{Kullie:2024,Hofmann:2019}. 
 This explains the already mentioned agreement between the $\tau_{\rm LC}$ 
 given by Hofmann (data-fit curve) and $\tau_{\rm dB}$ for small field 
 strengths.
 The agreement is satisfactory considering that the fitted LC-curve is 
 extended beyond the data range given in \cite{Hofmann:2019}.
 This leads us to the interaction time with the laser field in the 
 barrier area, since the LC-time according to Steinberg 
 \cite{Steinberg:1995,Spierings:2021} is related to the interaction 
 time within the barrier, which corresponds to the time spent in the 
 barrier or the dwell time $\tau_{\rm dwell}$ 
 \cite{Ramos:2020,Buettiker:1983}, which in turn equals $\tau_{\rm dB}, 
 \tau_{\rm barrier}$ according to the UTTP (and $\tau_{\rm dwell}$ of 
 Winful in Eq. \ref{Win}).

 The agreement shown in figure  \ref{figtdelt}(b) (see also Eq. \ref{tTdelt}) 
 suggests that the WM value (LC-time) and the interaction time within 
 the barrier region in the thick barrier limit, can be determined by 
 $\tau_{\rm dB}$ ($\approx \tau_{\rm dion}$ of  
 the nonadiabatic field calibration as given in Eq. \ref{tTdelt}). 
 We think this is similar to the measurement presented in 
 \cite{Ramos:2020} (measurement of the time spent in the barrier with  
 LC). 
 In addition, the back reaction of the measurement of the system can 
 be found from $\tau_{\rm T,i}$ as the following
  \begin{eqnarray*}
  \tau^{WM}_{backr}&=&
   \lim\limits_{F\ll F_{a}}\tau_{_{\rm T,i}}=
  \lim\limits_{F\ll F_{a}}\frac{1}{2}\frac{(I_{\rm p}-\delta_{z})}{4Z_{\rm eff}F} \\\nonumber 
 &=&\lim\limits_{F\ll F_{a}}\frac{1}{2}\frac{\varepsilon_{_{F}}}{4Z_{\rm eff}F}
 \approx\frac{1}{4 I_{\rm p}}= \lim\limits_{F\to 0}\tau_{_{\rm T,i}}
 \end{eqnarray*}
 where {$\varepsilon_{F}= (I_{\rm p}-\delta_{z})\sim 
 \nicefrac{{2 Z_{\rm eff}\, F}}{I_{\rm p}}$}
 is small under the WM condition (linear dependence on $F$), and we 
 used the expansion of {\scriptsize ${({I_{\rm p}^{2}-{4 Z_{\rm eff}F}}})^{1/2}$}
 for small $F$.
 
 However, the condition of WM is not necessary and we can assume that  
 $\tau_{_{\rm T,i}}$ always represents the back reaction of the system,
 which is generally consistent with the interpretation as the time 
 needed to reach the barrier entrance in the strong-field interaction, 
 see discussion after Eq. \ref{Tdi}.
 Finally, we note that $\tau_{\rm T,d},\tau_{_{\rm T,i}}$ can be 
 interpreted respectively, as forward and backward tunneling  
 \cite{Kullie:2015} and we can decompose $\tau_{_{\rm T,i}}$ similarly  
 as was done for Eq. \ref{TdF}, to   
 \begin{eqnarray}\nonumber
  \tau_{_{\rm T,d}}&=&\tau_{\rm dion}+\tau_{\rm dB} \mbox{ (forward tunneling)}\\
 \tau_{_{\rm T,i}}&=&\tau_{\rm dion}-\tau_{\rm dB}  \mbox{ (backward tunneling)}, 
\end{eqnarray}
 showing again that 
 \[\tau_{\rm dB}=\tau_{\rm dwell},  \]
 since, according to Steinberg \cite{Steinberg:1995}, we have 
 $\tau_{\rm dwell}=Re(\tau_{\rm T})=Re(\tau_{\rm R})$, where 
 $\tau_{\rm T}, \tau_{R}$ are the transmission and reflection 
 scattering channel times,  respectively. 
 They are the forward and backward (tunneling) scattering channels in our 
 model. 
 Of course, $\tau_{\rm dB}$ is real, as  we explained in our 
 previous work \cite{Kullie:2024}.
 In our forward and backward channels of adiabatic tunneling in Eq.  
 \ref{Tdi}, which is illustrated in figure \ref{figTI} (dashed-dotted, 
 red curve), the condition of a spatially symmetric barrier noted by 
 Steinberg \cite{Steinberg:1995} is reflected by the fact that the 
 horizontal channel happens along the (same) barrier width $d_{\rm B}=
 x_{\rm e,+}-x_{\rm e,-}$ for forward and  backward tunneling, compare 
 figure \ref{figTI}. 
 Finally we can write $\tau_{\rm T,{\rm (d,i)}}= \tau_{\rm dion}\pm\tau_{\rm dB}$ 
 and $\tau_{\rm dion} =(1/2)\, (\tau_{\rm T,d}+\tau_{_{\rm T,i}}), 
 \tau_{\rm dB} =(1/2)(\tau_{_{\rm T,d}}-\tau_{_{\rm T,i}})$, 
 where the symmetrization results in $\tau_{\rm dion}$ and the 
 anti-symmetrization results in the barrier time-delay or the dwell time.
 For further details and discussion, we would like to refer readers to 
 our previous work \cite{Kullie:2020}. 
 Forward and backward tunneling is equivalent to the transmission and 
 reflection of the wave packet in the traditional quantum tunneling 
 studies, which usually use numerical methods 
 \cite{Winful:2003,Lunardi:2019,Dumont:2023,Ma:2024,Carvalho:2002,Winful:2006II}, 
 whereas our model offers a simple tunneling approach with expressions 
 that agrees well the measurement result of the attoclock experiment.
\paragraph{\bf Conclusion and Outlook}
 The tunneling time has a universal behavior referred to as UTTP, 
 where the barrier (tunneling) time-delay can be defined by a simple 
 subtraction of adiabatic and nonadiabatic tunnel-ionization 
 time-delay, ($\tau_{\rm Ad}-\tau_{\rm dion})= \tau_{\rm dB}=  
 \tau_{\rm barrier}= \tau_{\rm dwell}$, see eqs  \ref{tdelt}, 
 (and \ref{Win} and \ref{Lun}).
 It is shown that $\tau_{\rm dB}$ agrees well with time-delay 
 $\tau_{\rm barrier,exp}$, which is obtained by fitting the 
 experimental data (see figure \ref{figtdelt}) as the difference 
 between the time-delays resulting from the adiabatic and nonadiabatic 
 field calibrations of the measurement of the attoclock 
 experiment, i.e. those of Landsman \cite{Landsman:2014II} and 
 Hofmann  \cite{Hofmann:2019} respectively. 
 More remarkably is that our result provides conceivable definitions of the tunnel-ionization   
 time-delay (adiabatic and nonadiabatic), the barrier (tunneling) 
 time-delay and the interpretation of the attoclock measurement.
 We also found that the barrier time-delay $\tau_{\rm dB}$ corresponds 
 to the LC-time and the interaction time 
 \cite{Buettiker:1983,Steinberg:1995,Ramos:2020,Spierings:2021}. 
 This is particularly evident in the limit of thick barrier, 
 where $\tau_{\rm dB}$ approaches the time-delay in the nonadiabatic 
 field calibration $\tau_{\rm dion}$, and where the back reaction is 
 small and the weak measurement approach is justified.
 We assume that there is a similarity to the measurement presented in 
 \cite{Ramos:2020}. 
 In the future, we will focus on faster-than-light tunneling or quantum  
 superluminal tunneling (QST), e.g. \cite{Dumont:2023}, one of the most 
 exciting phenomena in quantum physics.
 Our preliminary result shows that superluminality can be experimentally 
 investigated using the attoclock scheme and furthermore by the 
 numerical integration of the Schr\"odinger and Dirac equations, on 
 which we are currently working.
 Furthermore, it also shows that the condition on QST for the barrier 
 (tunneling) time-delay $\tau_{\rm dB}$ is less severe than expected 
 ($Z \sim 18$ for H-like atoms) and that the relativistic effects 
 are not crucial. 
 We think that the present work and future investigation on QST serve 
 as inspiration for further experimental and theoretical studies.
\subparagraph{Acknowledgments}
 I would like to thank Prof. Martin Garcia from the Theoretical
 Physics of the Institute of Physics at the University of Kassel
 for his kind support.
I am grateful to  A. S. Landsman and C. Hofmann 
for sending the data presented in \ref{figtdelt}. 
\providecommand{\newblock}{}

\end{document}